# Electronic Correlations and Hund's Rule Coupling in Trilayer Nickelate $La_4Ni_3O_{10}$

Zihao Huo[1], Peng Zhang[2,*], Zihan Zhang[1], Defang Duan[1,*], and Tian Cui[3,1]

[1]*Key Laboratory of Material Simulation Methods & Software of Ministry of Education, State Key Laboratory of Superhard Materials, College of Physics, Jilin University, Changchun 130012, China*

[2]*MOE Key Laboratory for Non-equilibrium Synthesis and Modulation of Condensed Matter, Shaanxi Province Key Laboratory of Advanced Functional Materials and Mesoscopic Physics, School of Physics, Xi'an Jiaotong University, Xi'an 710049, China*

[3]*Institute of High Pressure Physics, School of Physical Science and Technology, Ningbo University, Ningbo 315211, China*

*Corresponding author: duandf@jlu.edu.cn, zpantz@mail.xjtu.edu.cn

**Abstract:** Trilayer Ruddlesden–Popper phase $La_4Ni_3O_{10}$ has been observed with $T_c$ of ~ 30 K at high pressure in recent experiment, which further expanded the family of nickelate superconductors. In this study, we explored the effects of electronic correlations in $La_4Ni_3O_{10}$ using density function theory plus dynamical mean–field theory at ambient and high pressures. Our derived spectral functions and Fermi surface of ambient pressure phase are nicely consistent with the experimental results by angle–resolved photoemission spectroscopy, which emphasized the importance of electronic correlations in $La_4Ni_3O_{10}$. We also found the electronic correlations in pressured $La_4Ni_3O_{10}$ are both orbital–dependent and layer–dependent due to the presence of Hund's rule coupling. There is a competition between the Hund's rule coupling and the crystal–field splitting, and therefore the Ni–O layers with weaker crystal-field splitting energy would have stronger electronic correlations.

Exploration of high–$T_c$ superconductors has been one of the most important frontiers in condensed matter physics. As the first discovered high–$T_c$ superconductors, cuprates [1-4] have inspired the searches for superconductivity in oxide compounds. Due to the similarities with cuprates, nickelates have long been suggested as candidates of high–$T_c$ superconductors [5-8]. In 2019, a Sr–doped infinite–layer $NdNiO_2$ with $T_c$ of 15 K have been discovery [9]. Following this success, the superconductivity of infinite–layer nickelate have been widely explored [10-14]. Enormous investigations have been carried out concerning the physical properties of this superconducting system, such as the multiorbital nature [15], the Hund's metal behaviors [16], and the Kondo coupling [17-19]. Lately, superconductivity in bilayer Ruddlesden–Popper (RP) phase $La_3Ni_2O_7$ have been observed with a maximum $T_c$ of 80 K at pressure above 14 GPa [20-25]. As a second unconventional superconductor family with $T_c$ higher than the boiling temperature of liquid nitrogen, this breakthrough has attracted extensive attentions on the superconductivity mechanism and electronic correlations in $La_3Ni_2O_7$ [26-37], such as the orbital–selective behavior [33, 38], the Hund's rule coupling [30, 39, 40], and the interlayer superexchange interactions [41, 42]. The orbital–dependent electronic correlations and interlayer interactions have been experimental measured at ambient pressure [36, 37], which may support the interlayer superexchange and multiorbital nature.

Recently, trilayer RP phase $La_4Ni_3O_{10}$ have been observed with $T_c$ of ~ 30 K at pressure above 15 GPa [43-47], resulting from a structural transition from the $P2_1/a$ phase at the ambient pressure to the $I4/mmm$ phase at high pressure. Besides, some experiments reported that $La_4Ni_3O_{10}$ possess similar electronic state with $La_3Ni_2O_7$ and both nickelates have orbital–dependent electronic correlations at the ambient pressure [48, 49]. However, the $T_c$ of $La_4Ni_3O_{10}$ is much lower than $La_3Ni_2O_7$. The effective model of $La_4Ni_3O_{10}$ have been proposed [50-54] and suggested that the lower $T_c$ may due to the weaker hybridization between the Ni–$d_{z^2}$ and Ni–$d_{x^2-y^2}$ orbitals. Besides, using density function theory plus dynamical mean–field theory (DFT+DMFT), the researchers proposed that the electronic correlation effects at different Ni–O layer are not similar [55, 56]. Therefore, it is important to study the effects of electronic correlations in pressured $La_4Ni_3O_{10}$ and to explore the origination of layer-dependent electronic correlations.

In this work, we studied the effects of electronic correlations, especially the Hund's rule coupling, in trilayer nickelate $La_4Ni_3O_{10}$ using both the DFT+$U$ and the DFT+DMFT methods. We found that the electronic structures of $La_4Ni_3O_{10}$ in $P2_1/a$ phase being calculated by DFT+DMFT are nicely consistent with the experimental observations, therefore emphasizing the importance of multi-orbital electronic correlations in modelling this trilayer nickelate. We then studied the high pressure $I4/mmm$ phase of $La_4Ni_3O_{10}$ by DFT+DMFT, and found that the electronic correlations of Ni-$d_{z^2}$ orbitals are stronger than these of Ni-$d_{x^2-y^2}$ orbitals, presenting an orbital–dependent nature. Besides, the magnitudes of electronic correlations are dependent on the Ni–O layers. Both the orbital–dependent and the layer–dependent features of electronic correlations in $La_4Ni_3O_{10}$ are originated from the Hund's rule coupling. By analyzing the evolution of atomic spin state probability and Hund's rule coupling, we suggested that a competition between the Hund's rule coupling and the crystal-field splitting. The Ni–O layer with weaker crystal-field splitting energy would have a stronger correlation.

$La_4Ni_3O_{10}$ has a monoclinic $P2_1/a$ symmetry at the ambient pressure, which can be termed as an inter–growth of three $NiO_6$ octahedra planes and two La–O fluorite–type layers stacking along the *c* directions. Adopting the crystal structure from previous experiments [43], we first calculated its band structures and density of states (DOS) using the DFT+*U* method as implemented in the Wien2k package [57] and shown in Fig. 1(a). An effective $U$ = 3.5 eV is employed in the correlated Ni-3*d* orbitals, in line with previous bilayer and trilayer nickelate DFT calculations [37, 53]. The details of computational method are provided in the Supplemental Materials. Our results show that both the Ni–$d_{z^2}$ and Ni–$d_{x^2-y^2}$ orbitals cross the Fermi level ($E_F$), which is different from bilayer *Amam* phase of $La_3Ni_2O_7$ at the ambient pressure with only Ni–$d_{x^2-y^2}$ orbitals cross the $E_F$ [37]. There are two pockets around the Γ point in Fig. 1(c), including a smaller electronic $\gamma$ pocket and a relatively larger electronic $\alpha$ pocket. Although the derived Fermi surface by DFT is similar to those reported in previous DFT calculations [44, 48], they are contradicted with the angle–resolved photoemission spectroscopy (ARPES) results [44, 48]. The Fermi surface from ARPES has only one pocket around the Γ point, as shown in the background in shaded grey.

We then calculated the ***k***–resolved spectral functions, the DOS and the Fermi surface of $La_4Ni_3O_{10}$ in $P2_1/a$ phase at 100 K using the DFT+DMFT package [58] with $U$ = 3.5 eV and $J$ = 1.0 eV (Fig. 1(b, d)). Compared with the DFT result, both the $\alpha$, $\beta$, and $\gamma$ bands are strongly renormalized and become very flat near the $E_F$. The energy range of Ni–$d_{z^2}$ and Ni–$d_{x^2-y^2}$ DOS are only about 0.2 eV, a factor of about 3.5 times narrower than that from the DFT result. Besides, for energy below -0.2 eV, the spectral functions become more incoherent due to the strong electronic interaction, which is also consistent with the ARPES results. It is noticed that the $\gamma$ band does not cross the $E_F$ anymore, therefore the $\gamma$ pocket vanished after inclusion of the electron-electron scattering and the multi–orbital coupling by DMFT. In Fig. 1(d), there is only one electronic α pocket in octagon shape around the Γ point, and the Fermi surface by DFT+DMFT nicely overlaps with that from APRES. Both the spectral functions and the derived Fermi surface indicate that DFT+DMFT method can provide improved description of $La_4Ni_3O_{10}$ than DFT through benchmarks with the APRES experimental results. The consistency between our DFT+DMFT calculations and the ARPES results thus emphasized the importance of multi-orbital electronic correlations in $La_4Ni_3O_{10}$, and justified our choice of the Hubbard *U* and the Hund's rule coupling *J* in this work.

Next, we studied the $La_4Ni_3O_{10}$ at high pressure of 30.5 GPa by DFT+DMFT, which adopts a tetragonal symmetry (space group: *I*4/*mmm*) (Fig. 2(a)) with lattice parameters a = 3.709 Å and c = 26.716 Å [43]. We define the upper and lower Ni–O layers as the outer layers and the middle Ni–O layers as the inner layers. Fig. 2(b–e) represent the DOS and the hybridization functions of Ni–$d_{z^2}$ and Ni–$d_{x^2-y^2}$ orbitals at 100 K with $U$ = 3.5 eV and $J$ = 1 eV. For the inner layer, we can see that both the DOS of Ni–$d_{z^2}$ and Ni–$d_{x^2-y^2}$ orbitals formed a peak near the $E_F$. It suggests that Ni–$d_{z^2}$ and Ni–$d_{x^2-y^2}$ orbitals producing characteristic quasiparticle spectra peaks in the inner layers. The existence of high hybridization peaks of Ni–$d_{z^2}$ and Ni–$d_{x^2-y^2}$ orbitals at around 0.3 eV in Fig. 2(d) indicated that these two orbitals have strong hybridization with the O–$p$ orbitals. The strength of hybridization peak of Ni–$d_{z^2}$ and Ni–$d_{x^2-y^2}$ orbitals in the outer layers at around 0.3 eV is lower than the inner layer, which indicate that the hybridization strength of these two orbitals in outer layer is weaker than these in inner layers.

The imaginary parts of the Matsubara–frequency self–energy functions are analyzed to unveil the effects of electronic correlations in Ni–$d_{z^2}$ and Ni–$d_{x^2-y^2}$ orbitals. As shown in Fig. 3(a), the imaginary part of self–energy functions $\mathrm{Im}\sum(i\omega_n)$ in both the inner and the outer layers extrapolate to zero as $\omega \to 0$, which suggests the coherent electron–scattering in $La_4Ni_3O_{10}$ at 100 K. In both the inner and the outer layers, the self–energy function magnitudes of Ni–$d_{z^2}$ orbitals are larger than those of Ni–$d_{x^2-y^2}$ orbitals, which indicates the quasi–particle weight $Z = \left[1 - \frac{\partial \,\mathrm{Im}\sum(\omega_n)}{\partial \omega_n}\right]^{-1}\Big|_{\omega_n \to 0^+}$ of the Ni–$d_{z^2}$ orbitals would be smaller. Since the quasiparticle weight $Z$ is inversely proportional to the strength of electronic correlations, the correlation strength of the Ni–$d_{z^2}$ orbitals would be larger than that of the Ni–$d_{x^2-y^2}$ orbitals. Our discovery of the orbital–dependent electronic correlation feature and the stronger correlated Ni–$d_{z^2}$ orbitals in $I4/mmm$ phase of $La_4Ni_3O_{10}$ is consistent with previous experimental observation at ambient pressure[48]. Similarly, the self–energy magnitudes of Ni–$d_{z^2}$ and Ni–$d_{x^2-y^2}$ orbitals in outer layers are always larger than these in inner layers, which suggest the correlation effects of the outer layer is stronger than these of the inner layers. Therefore, the $I4/mmm$ phase of $La_4Ni_3O_{10}$ presents a layer-dependent electronic correlations feature.

Since the Hund's rule coupling $J$ have been suggested important in pressured $La_3Ni_2O_7$ [30], we examined the effects of Hund's rule coupling by reducing it to 0 eV. The imaginary parts of the Matsubara–frequency self–energy functions with $U$ = 3.5 eV and $J$ = 0 eV have been plotted in Fig. 3(b). Either the orbital–dependent diversity or the layer–dependent diversity has been strongly quenched in comparison with the self-energies at $U = 3.5$ eV and $J = 1$ eV in Fig. 1(a), which suggests the orbital–dependent and the layer–dependent electronic correlation is mainly due to the Hund's rule coupling.

The effective mass enhancement $m^*/m = Z^{-1}$ as functions of $U$ and $J$ at 100 K are plotted in Fig. 3 (c–d). To avoid a large error bar in the analytic continuation, we directly calculate the quasi–particle weight $Z$ from the fourth–order polynomial fit to the self–energies at the first ten Matsubara frequencies. For $U = 3.5$ eV and $J = 0$ eV, the $m^*/m$ of Ni–$d_{z^2}$ and Ni–$d_{x^2-y^2}$ orbitals in both inner and outer layers are similar. And when $J$ increased from 0 to 1 eV, the $m^*/m$ of Ni–$d_{z^2}$ and Ni–$d_{x^2-y^2}$ orbitals in the inner layers increases from 1.28 and 1.27 to 1.82 and 1.75, and the increasing ratio are 1.42 and 1.38, respectively. In contrast, $m^*/m$ of the Ni–$d_{z^2}$ and Ni–$d_{x^2-y^2}$ orbitals in the outer layers increases from 1.34 and 1.3 to 2.26 and 1.88, and the increasing ratio are 1.69 and 1.45, respectively. The magnitude of electronic correlation of Ni–$d$ electrons in the outer layers are strongly depend on the Hund's rule coupling than these in the inner layers. Besides, the difference of $m^*/m$ between the Ni–$d_{z^2}$ and Ni–$d_{x^2-y^2}$ orbitals in the outer layers are higher than that in the inner layers, indicating a stronger orbital-dependence of correlations in the outer layers. When $J = 0$ eV and $U$ increased from 2 to 6 eV, the $m^*/m$ of Ni–$d_{z^2}$ and Ni–$d_{x^2-y^2}$ orbitals in the inner layers increases from 1.13 and 1.12 to 1.51 and 1.48, and the increasing ratio are 1.34 and 1.32, respectively. The $m^*/m$ of Ni–$d_{z^2}$ and Ni–$d_{x^2-y^2}$ orbitals in the outer layers increases from 1.15 and 1.13 to 1.6 and 1.55, and the increasing ratio are 1.39 and 1.37, respectively. Therefore, the evolution of $m^*/m$ for for Ni–$d_{z^2}$ and Ni–$d_{x^2-y^2}$ orbitals in both the inner and the outer layers indicate a weak dependence on the Hubbard $U$. We suggest that in $I4/mmm$ phase of $La_4Ni_3O_{10}$ the Hund's rule coupling plays more important role in modulating the strength of electronic correlations than the Hubbard $U$.

Fig. 4(a) shows the evolution of instantaneous local magnetic moment $M$ as a function of temperature. The local magnetic moment is derived by $M = 2\sqrt{\sum_k P_k S_k^{z2}}$, where $P_k$ is the probability of atomic states. The local magnetic moment $M$ in both the inner and the outer layers slightly increases when the temperature increased. And this phenomenon may be incurred by the antiferromagnetic spin correlations [59], which arise from the interlayer superexchange coupling between two Ni–$d_{z^2}$ spins through the apical O–$p_z$ orbital. Besides, the local magnetic moment $M$ at $J = 1$ eV in both the inner and the outer layers are larger than those at $J = 0$ eV. Therefore, the Hund's rule coupling would induce extra high-spin states in the $I4/mmm$ phase of $La_4Ni_3O_{10}$.

We also calculated the atomic spin state probability for the electron occupation number $N = 7$ and 8 as a function of $J$ in Fig. 4 (c–d). For $J = 0$ eV, the maximum possibility atomic spin state is low–spin states with $N = 8$, $S = 0$ in both the inner and the outer layers. It is worth to note that with the increment of $J$, the possibility of low–spin states with $N = 8$, $S = 0$ decreased and the high–spin states with $N = 8$, $S = 1$ increased. By fitting the data, we found that when $J$ is higher than 1.02 eV, the maximum possibility atomic spin state in inner layers is the high–spin state with $N = 8$, $S = 1$. For the outer layers, this transition starts when $J$ is higher than 0.82 eV. Therefore, there is a competition between the high–spin states and the low-spin states in $I4/mmm$ phase of $La_4Ni_3O_{10}$, which has also been suggested in infinite–layer nickelate [15, 60]. Since the atomic spin states are mainly controlled by the crystal-field splitting if the Hund's rule coupling is not considered, the partition ratio between the high–spin states and the low–spin states may originate from the competition between Hund's rule coupling and the crystal-field splitting. And the high–spin states induced by Hund's rule coupling would cause the orbital blocking effect [61], resulting an enhancement of the electronic correlation in $I4/mmm$ phase of $La_4Ni_3O_{10}$. Besides, the transition value of $J$ in inner layers are much larger than that in outer layers, which suggests a large crystal-field splitting energy and a more itinerant nature in inner layers. We also calculated the atomic spin state probability for the electron occupation number $N = 7$ and $N = 8$ as a function of Hund's rule coupling $J$ for $I4/mmm$ phase of $La_3Ni_2O_7$ at 30 GPa (the crystal structure was adopted from previous research [34]) and plotted the results in Fig. 4 (b). A similar phenomenon has also been found in $I4/mmm$ phase of $La_3Ni_2O_7$ with a transition value of $J = 0.77$ eV, which is similar with that of outer layers and much lower than that of inner layers of $I4/mmm$ phase of $La_4Ni_3O_{10}$. Since the hybridization between the Ni–$d_{z^2}$ and Ni–$d_{x^2-y^2}$ orbitals has been suggested to be important in the superconductivity in both $La_3Ni_2O_7$ and $La_4Ni_3O_{10}$ [42, 51, 62], a weak strength of Hund's rule coupling in inner layers may not conducive for superconductivity and resulting a low $T_c$ in pressured $La_4Ni_3O_{10}$.

Table I shows the local occupancy numbers $n_d$, and the mass enhancement $m^*/m$ of Ni–$d_{z^2}$ and Ni–$d_{x^2-y^2}$ orbitals for $P2_1/a$ phase of $La_4Ni_3O_{10}$, $I4/mmm$ phase of $La_4Ni_3O_{10}$, and $I4/mmm$ phase of $La_3Ni_2O_7$. For $P2_1/a$ phase of $La_4Ni_3O_{10}$ at ambient pressure, the $n_d$ of Ni–$d_{z^2}$ orbitals in inner layer is 1.167, which is large than that of Ni–$d_{x^2-y^2}$ orbitals (1.015). While in outer layers, the $n_d$ of these two orbitals are similar. It suggests that the Ni–$d_{x^2-y^2}$ orbitals in inner layers have more hole doping than that of Ni–$d_{z^2}$ orbitals. Besides, the $m^*/m$ of Ni–$d_{z^2}$ orbitals in both inner and outer layers are larger than that of Ni–$d_{x^2-y^2}$ orbitals, resulting a more localized behavior of Ni–$d_{z^2}$ orbitals and being consistent with the experimental results [44, 48]. For $I4/mmm$ phase of $La_4Ni_3O_{10}$ at 30.5 GPa, the $n_d$ of both Ni–$d_{z^2}$ and Ni–$d_{x^2-y^2}$ orbitals is lower than the $P2_1/a$ phase. These results show that $La_4Ni_3O_{10}$ at high pressure has more hole doping compared with it at the ambient pressure, which is corresponding with the lower $m^*/m$ value and the more itinerated nature of these two orbitals in *I*4/*mmm* phase of $La_4Ni_3O_{10}$ at high pressure. And this phenomenon is similar with the Sr-doped infinite nickelate [16]. The electronic correlation strength of Ni–$d_{z^2}$ and Ni–$d_{x^2-y^2}$ orbitals in both inner and outer layers of *I*4/*mmm* phase of $La_4Ni_3O_{10}$ are weaker than these orbitals in *I*4/*mmm* phase of $La_3Ni_2O_7$, especially for the inner layers. It seems suggest that the Ni–$e_g$ orbitals with more itinerant nature are not beneficial for the superconductivity in the RP phase of nickelate.

In summary, we performed comprehensive study on the effects of electronic correlations in $La_4Ni_3O_{10}$ at both the ambient pressure and high pressure using both DFT+U and DFT+DMFT methods. Our calculated spectral functions and Fermi surface of the $P2_1/a$ phase of $La_4Ni_3O_{10}$ are nicely consistent with ARPES experimental results. By analyzing the self–energy functions and the quasi–particle weight, we found that the orbital–dependent and the layer–dependent feature in *I*4/*mmm* phase of $La_4Ni_3O_{10}$ mainly originate from the Hund's rule coupling. Moreover, there existed a competition between the Hund's rule coupling and the crystal field splitting, which controls the partitioning between the high-spin and the low–spin states in *I*4/*mmm* phase of $La_4Ni_3O_{10}$. The layer–dependent feature in *I*4/*mmm* phase of $La_4Ni_3O_{10}$ is caused by the different crystal-field splitting energy. Out study emphasize the importance of multi–orbital electronic correlations in trilayer nickelate $La_4Ni_3O_{10}$ and can help understanding experimental results of this fancy superconducting compound.

This work was supported by National Natural Science Foundation of China (Grants Nos. 12122405, 52072188 and 12274169), National Key R&D Program of China (Grants Nos. 2022YFA1402304 and 2023YFA1406200), Fundamental Research Funds for the Central Universities (Grants No. xzy022023011 and xhj032021014-04), and Program for Science and Technology Innovation Team in Zhejiang (Grant No. 2021R01004). Some of the calculations were performed at the High Performance Computing Center of Jilin University and using TianHe-1(A) at the National Supercomputer Center in Tianjin.

See Supplemental Material for ***k***–resolved spectral functions, DOS, Fermi surface of $P2_1/a$ phase of $La_4Ni_3O_{10}$ for several different Hubbard $U$ and Hund's rule coupling $J$, DOS of *I*4/*mmm* phase of $La_4Ni_3O_{10}$, structure information of RP phase $La_3Ni_2O_7$ and $La_4Ni_3O_{10}$, and computational details.

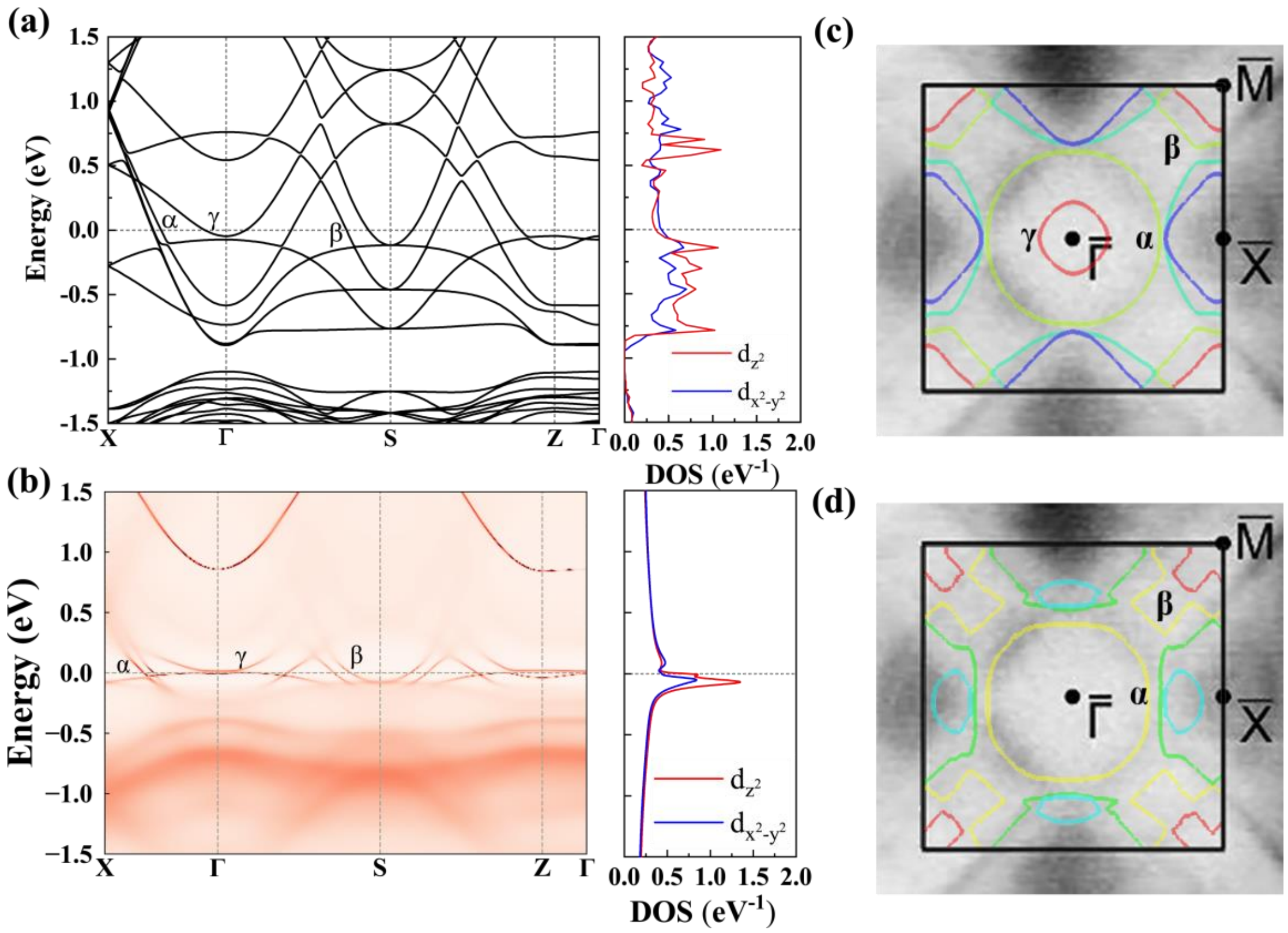

FIG 1. Electronic structures of $La_4Ni_3O_{10}$ in $P2_1/a$ phase. (a) Electronic band structures and DOS from DFT+$U$ ($U$ = 3.5 eV). (b) ***k***–resolved spectral functions and DOS from DFT+DMFT at 100 K. The horizontal gray dashed line represents the Fermi level. The colored Fermi surfaces being calculated by (c) DFT+$U$ and (d) DFT+DMFT at 100 K. Different colors represent different band index. The shaded Fermi surfaces in the background are recreated from Zhang *et al*.' experiment [44].

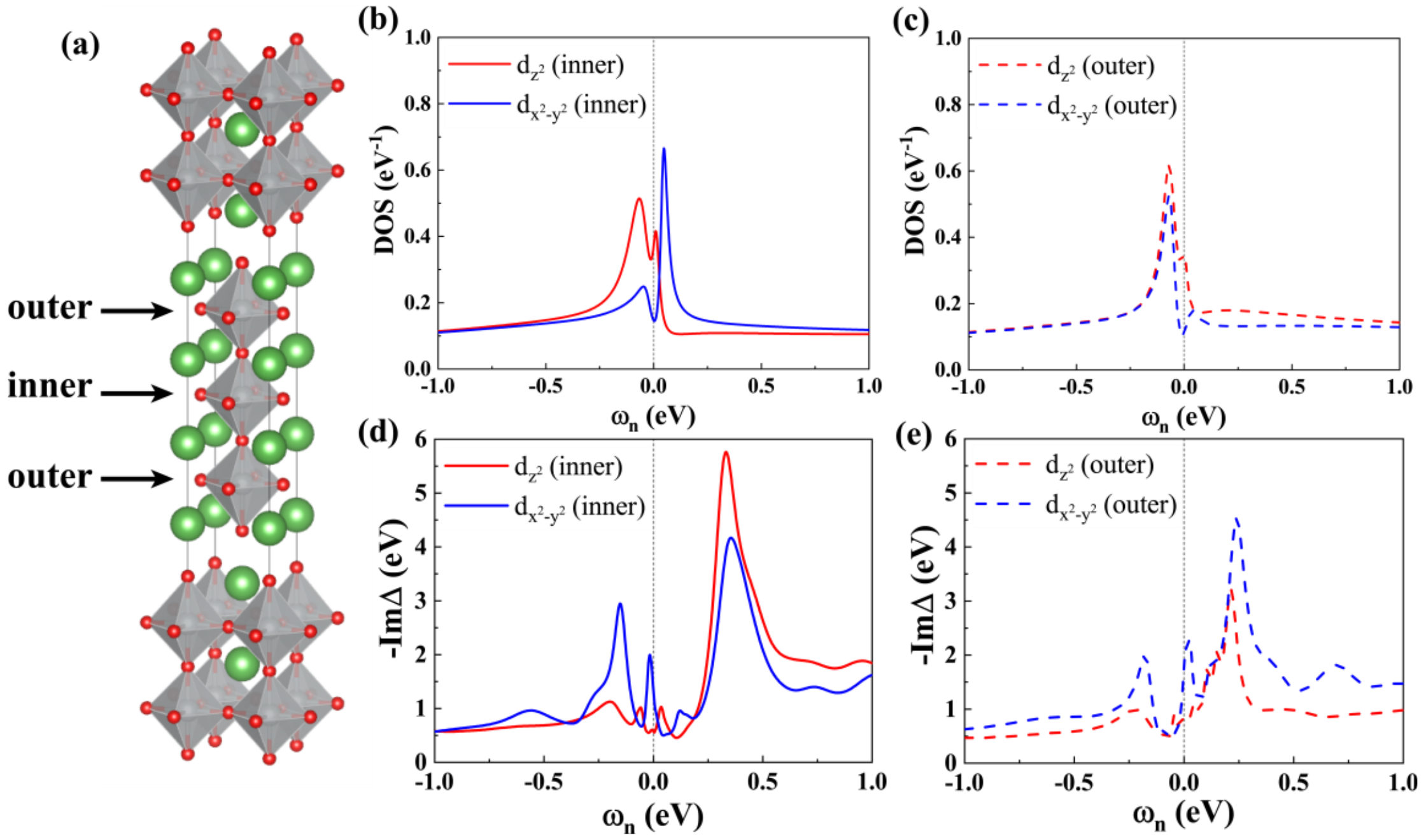

FIG. 2. (a) Crystal structures of the *I*4/*mmm* phase of $La_4Ni_3O_{10}$. The green, grey, and red spheres represent La, Ni, and O atoms, respectively. Orbital–dependent electronic DOS of Ni in (b) inner and (c) outer layer from DFT+DMFT at 100 K. The grey dashed line indicated the Fermi level. (d–e) Negative of the imaginary part of hybridization function in inner and outer layers from DFT+DMFT at 100 K, respectively.

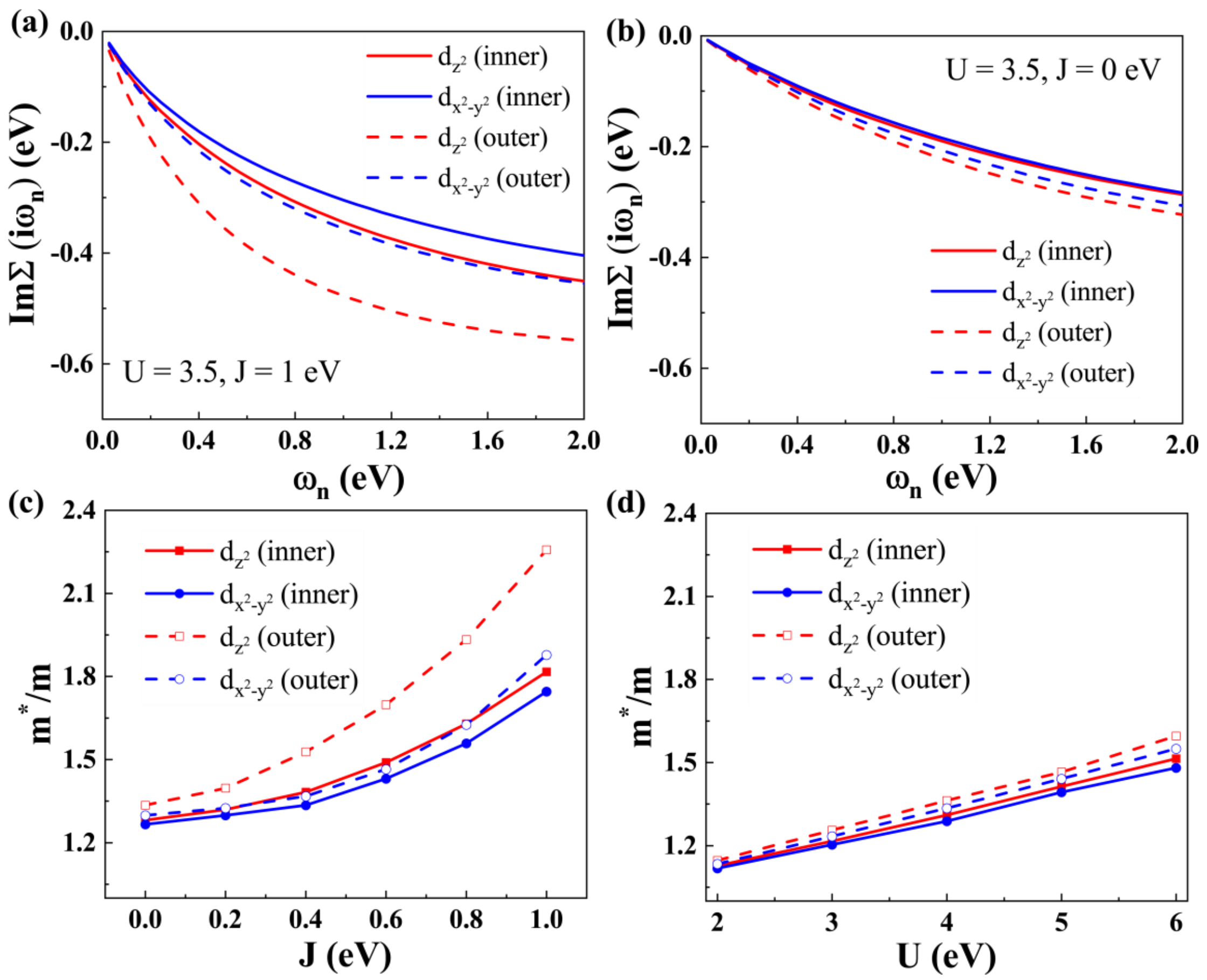

FIG 3. Imaginary parts of the orbital-dependent self-energy functions ImΣ ($i\omega_n$) for (a) $U$ = 3.5 eV, $J$ = 1 eV and (b) $U$ = 3.5 eV, $J$ = 0 eV on the Mastsubara axis from DFT+DMFT at 100 K. The effective mass enhancement $m^*/m$ of Ni–$d_{z^2}$ and Ni–$d_{x^2-y^2}$ orbitals in inner and outer layers from DFT+DMFT at 100 K as functions of (c) Hund's rule coupling $J$ at $U$ = 3.5 eV and (d) Hubbard $U$ at $J$ = 0 eV.

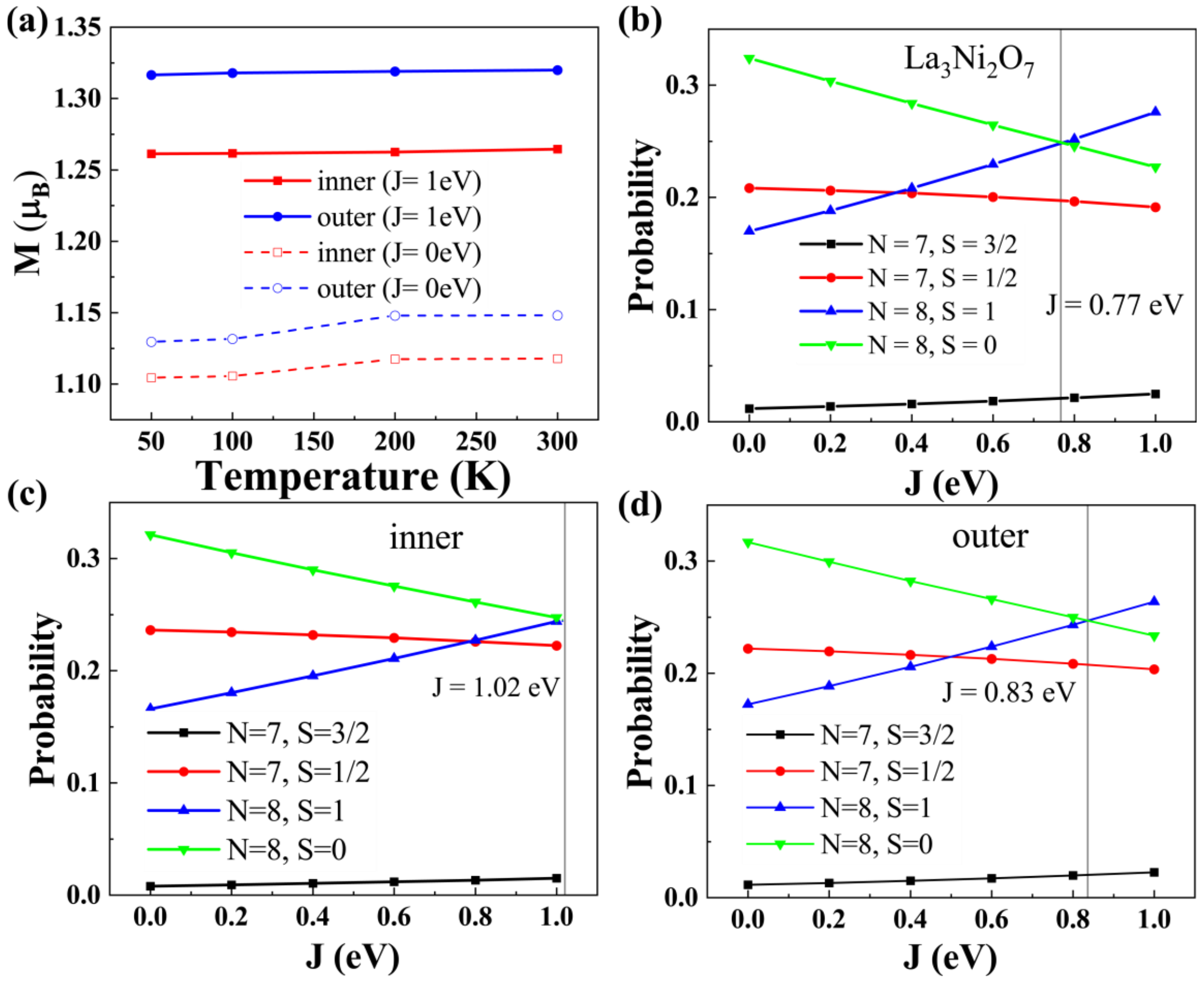

FIG 4. (a) Evolution of instantaneous local magnetic moment $M$ as a function of the temperature with $U = 3.5$ eV. Calculated probability for the electron occupation number $N = 7$ and $N = 8$ as a function of Hund's rule coupling $J$ for (b) $I4/mmm$ phase of $La_3Ni_2O_7$, (c) inner layers, and (d) outer layers from DFT+DMFT at 100 K.

TABLE I. The electronic occupancy numbers $n_d$, and the mass enhancement $m^*/m$ of Ni–$d_{z^2}$ and Ni–$d_{x^2-y^2}$ orbitals for $P2_1/a$ phase of $La_4Ni_3O_{10}$, $I4/mmm$ phase of $La_4Ni_3O_{10}$, and $I4/mmm$ phase of $La_3Ni_2O_7$ from DFT+DMFT at 100 K. The i for inner layers and o for outer layers.

| | $P2_1/a$–$La_4Ni_3O_{10}$ @0 GPa | I4/*mmm*–$La_4Ni_3O_{10}$ @30.5 GPa | I4/*mmm*–$La_3Ni_2O_7$ @30 GPa |
|---|---|---|---|
| $n_d$–$d_{z^2}$ | 1.167 (i)<br>1.112 (o) | 1.034 (i)<br>1.079 (o) | 1.128 |
| $n_d$–$d_{x^2-y^2}$ | 1.015 (i)<br>1.116 (o) | 0.996 (i)<br>1.045 (o) | 1.044 |
| $m^*/m$–$d_{z^2}$ | 2.827 (i)<br>3.378 (o) | 1.817 (i)<br>2.257 (o) | 2.667 |
| $m^*/m$–$d_{x^2-y^2}$ | 2.601 (i)<br>2.968 (o) | 1.745 (i)<br>1.878 (o) | 2.052 |

## References

[1] J. Bednorz and K. Müller, Possible high $T_c$ Superconductivity in the Ba-La-Cu-O system, Z. Phys. B Condens. Matter **64**, 189 (1986).

[2] M. K. Wu, J. R. Ashburn, C. J. Torng, P. H. Hor, R. L. Meng, L. Gao, Z. J. Huang, Y. Q. Wang and C. W. Chu, Superconductivity at 93 K in a New Mixed-Phase Y-Ba-Cu-O Compound System at Ambient Pressure, Phys. Rev. Lett. **58**, 908 (1987).

[3] P. Lee, N. Nagaosa and X. Wen, Doping a Mott insulator: Physics of high-temperature superconductivity, Rev. Mod. Phys.**78**, 17 (2006).

[4] B. Keimer, S. Kivelson, M. Norman, S. Uchida and J. Zaanen, From quantum matter to high-temperature superconductivity in copper oxides, Nature **518**, 179 (2015).

[5] V. I. Anisimov, D. Bukhvalov and T. M. Rice, Electronic structure of possible nickelate analogs to the cuprates, Phys. Rev. B **59**, 7901 (1999).

[6] K. W. Lee and W. E. Pickett, Infinite-layer $LaNiO_2$: $Ni^{1+}$ is not $Cu^{2+}$, Phys. Rev. B **70**, 165109 (2004).

[7] J. Chaloupka and G. Khaliullin, Orbital Order and Possible Superconductivity in $LaNiO_3/LaMO_3$ Superlattices, Phys. Rev. Lett. **100**, 016404 (2008).

[8] J. Zhang, Y. Chen, D. Phelan, H. Zheng, M. Norman and J. Mitchell, Stacked charge stripes in the quasi-2D trilayer nickelate $La_4Ni_3O_8$, Proc. Natl. Acad. Sci. U.S.A. **113**, 8945 (2016).

[9] D. Li, K. Lee, B. Wang, M. Osada, S. Crossley, H. Lee, Y. Cui, Y. Hikita and H. Hwang, Superconductivity in an infinite-layer nickelate, Nature **572**, 624 (2019).

[10] M. Osada, B. Wang, B. Goodge, S. Harvey, K. Lee, D. Li, L. Kourkoutis and H. Hwang, Nickelate Superconductivity without Rare-Earth Magnetism: $(La,Sr)NiO_2$, Adv. Mater. **33**, 2104083 (2021).

[11] N. Wang, M. Yang, Z. Yang, K. Chen, H. Zhang, Q. Zhang, Z. Zhu, Y. Uwatoko, L. Gu, X. Dong, J. Sun, K. Jin and J. Cheng, Pressure-induced monotonic enhancement of $T_c$ to over 30 K in superconducting $Pr_{0.82}Sr_{0.18}NiO_2$ thin films, Nat. Commun. **13**, 4367 (2022).

[12] W. Sun, Y. Li, R. Liu, J. Yang, J. Li, W. Wei, G. Jin, S. Yan, H. Sun, W. Guo, Z. Gu, Z. Zhu, Y. Sun, Z. Shi, Y. Deng, X. Wang and Y. Nie, Evidence for Anisotropic Superconductivity Beyond Pauli Limit in Infinite-Layer Lanthanum Nickelates, Adv. Mater. **35**, 2303400 (2023).

[13] D. Li, B. Y. Wang, K. Lee, S. P. Harvey, M. Osada, B. H. Goodge, L. F. Kourkoutis and H. Y. Hwang, Superconducting Dome in $Nd_{1-x}Sr_xNiO_2$ Infinite Layer Films, Phys. Rev. Lett. **125**, 027001 (2020).

[14] S. Zeng, C. Tang, X. Yin, C. Li, M. Li, Z. Huang, J. Hu, W. Liu, G. Omar, H. Jani, Z. Lim, K. Han, D. Wan, P. Yang, S. Pennycook, A. Wee and A. Ariando, Phase Diagram and Superconducting Dome of Infinite-Layer $Nd_{1-x}Sr_xNiO_2$ Thin Films, Phys. Rev. Lett. **125**, 147003 (2020).

[15] X. Wan, V. Ivanov, G. Resta, I. Leonov and S. Y. Savrasov, Exchange interactions and sensitivity of the Ni two-hole spin state to Hund's coupling in doped $NdNiO_2$, Phys. Rev. B **103**, 075123 (2021).

[16] Y. Wang, C. J. Kang, H. Miao and G. Kotliar, Hund's metal physics: From $SrNiO_2$ to $LaNiO_2$, Phys. Rev. B **102**, 161118(R) (2020).
[17] G. M. Zhang, Y. F. Yang and F. C. Zhang, Self-doped Mott insulator for parent compounds of nickelate superconductors, Phys. Rev. B **101**, 020501(R) (2020).
[18] M. Hepting, D. Li, C. Jia, H. Lu, E. Paris, Y. Tseng, X. Feng, M. Osada, E. Been, Y. Hikita, Y. Chuang, Z. Hussain, K. Zhou, A. Nag, M. Garcia-Fernandez, M. Rossi, H. Huang, D. Huang, Z. Shen, T. Schmitt, H. Hwang, B. Moritz, J. Zaanen, T. Devereaux and W. Lee, Electronic structure of the parent compound of superconducting infinite-layer nickelates, Nat. Mater. **19**, 381 (2020).
[19] B. Goodge, D. Li, K. Lee, M. Osada, B. Wang, G. Sawatzky, H. Hwang and L. Kourkoutis, Doping evolution of the Mott-Hubbard landscape in infinite-layer nickelates, Proc. Natl. Acad. Sci. U.S.A. **118**, e2007683118 (2021).
[20] H. Sun, M. Huo, X. Hu, J. Li, Z. Liu, Y. Han, L. Tang, Z. Mao, P. Yang, B. Wang, J. Cheng, D. Yao, G. Zhang and M. Wang, Signatures of superconductivity near 80 K in a nickelate under high pressure, Nature **621**, 493 (2023).
[21] J. Hou, P. Yang, Z. Liu, J. Li, P. Shan, L. Ma, G. Wang, N. Wang, H. Guo, J. Sun, Y. Uwatoko, M. Wang, G. Zhang, B. Wang and J. Cheng, Emergence of High-Temperature Superconducting Phase in Pressurized $La_3Ni_2O_7$, Chin. Phys. Lett. **40**, 117302 (2023).
[22] Y. Zhang, D. Su, Y. Huang, H. Sun, M. Huo, Z. Shan, K. Ye, Z. Yang, R. Li, M. Smidman, M. Wang, L. Jiao and H. Yuan, High-temperature supercondcutivity with zero-resistance and strange metal behavior in $La_3Ni_2O_{7-\delta}$, Nat. Phys., https://doi.org/10.1038/s41567-024-02515-y (2024).
[23] Y. Zhou, J. Guo, S. Cai, H. Sun, P. Wang, J. Zhao, J. Han, X. Chen, Q. Wu, Y. Ding, M. Wang, T. Xiang, H. Mao and L. Sun, Investigations of key issues on the reproducibility of high-$T_c$ superconductivity emerging from compressed $La_3Ni_2O_7$, arXiv preprint arXiv:2311.12361 (2023).
[24] G. Wang, N. N. Wang, X. L. Shen, J. Hou, L. Ma, L. F. Shi, Z. A. Ren, Y. D. Gu, H. M. Ma, P. T. Yang, Z. Y. Liu, H. Z. Guo, J. P. Sun, G. M. Zhang, S. Calder, J. Q. Yan, B. S. Wang, Y. Uwatoko and J. G. Cheng, Pressure-Induced Superconductivity In Polycrystalline $La_3Ni_2O_{7-\delta}$, Phys. Rev. X **14**, 011040 (2024).
[25] J. Li, P. Ma, H. Zhang, X. Huang, C. Huang, M. Huo, D. Hu, Z. Dong, C. He, J. Liao, X. Chen, T. Xie, H. Sun and M. Wang, arXiv preprint arXiv:2404.11369 (2024).
[26] Z. Luo, X. Hu, M. Wang, W. Wú and D. X. Yao, Bilayer Two-Orbital Model of $La_3Ni_2O_7$ under Pressure, Phys. Rev. Lett. **131**, 126001 (2023).
[27] X. Z. Qu, D. W. Qu, J. Chen, C. Wu, F. Yang, W. Li and G. Su, Bilayer $t$-$J$-$J_\perp$ Model and Magnetically Mediated Pairing in the Pressurized Nickelate $La_3Ni_2O_7$, Phys. Rev. Lett. **132**, 036502 (2024).
[28] D. Shilenko and I. Leonov, Correlated electronic structure, orbital-selective behavior, and magnetic correlations in double-layer $La_3Ni_2O_7$ under pressure, Phys. Rev. B **108**, 125105 (2023).
[29] W. Wú, Z. Luo, D. Yao and M. Wang, Superexchange and charge transfer in the nickelate superconductor $La_3Ni_2O_7$ under pressure, Sci. China-Phys. Mech. **67**, 117402 (2024).

[30] Y. Cao and Y. F. Yang, Flat bands promoted by Hund's rule coupling in the candidate double-layer high-temperature superconductor $La_3Ni_2O_7$ under high pressure, Phys. Rev. B **109**, L081105 (2024).
[31] V. Christiansson, F. Petocchi and P. Werner, Correlated Electronic Structure of $La_3Ni_2O_7$ under Pressure, Phys. Rev. Lett. **131**, 206501 (2023).
[32] F. Lechermann, J. Gondolf, S. Bötzel and I. M. Eremin, Electronic correlations and superconducting instability in $La_3Ni_2O_7$ under high pressure, Phys. Rev. B **108**, L201121 (2023).
[33] Y. Zhang, L. Lin, A. Moreo and E. Dagotto, Electronic structure, dimer physics, orbital-selective behavior, and magnetic tendencies in the bilayer nickelate superconductor $La_3Ni_2O_7$ under pressure, Phys. Rev. B **108**, L180510 (2023).
[34] Z. Huo, Zhihui Luo, P. Zhang, A. Yang, Z. Liu, X. Tao, Z. Zhang, S. Guo, Q. Jiang, W. Chen, D. Yao, D. Duan and T. Cui, Modulation of the Octahedral Structure and Potential Superconductivity of $La_3Ni_2O_7$ through Strain Engineering, arXiv preprint arXiv:2404.11001 (2024).
[35] Y. Liu, J. Mei, F. Ye, W. Chen and F. Yang, $s^{\pm}$-Wave Pairing and the Destructive Role of Apical-Oxygen Deficencies in $La_3Ni_2O_7$ under Pressure, Phys. Rev. Lett. **131**, 236002 (2023).
[36] X. Chen, J. Choi, Z. Jiang, J. Mei, K. Jiang, J. Li, S. Agrestini, M. Garcia-Fernandez, X. Huang, H. Sun, D. Shen, M. Wang, J. Hu, Y. Lu, K. Zhou and D. Feng, Electronic and magnetic excitations in $La_3Ni_2O_7$, arXiv preprint arXiv:2401.12657 (2024).
[37] J. Yang, H. Sun, X. Hu, Y. Xie, T. Miao, H. Luo, H. Chen, B. Liang, W. Zhu, G. Qu, C. Chen, M. Huo, Y. Huang, S. Zhang, F. Zhang, F. Yang, Z. Wang, Q. Peng, H. Mao, G. Liu, Z. Xu, T. Qian, D. Yao, M. Wang, L. Zhao and X. Zhou, Orbital-dependent electron correlation in double-layer nickelate $La_3Ni_2O_7$, Nat. Commun. **15**, 4373 (2024).
[38] J. Chen, F. Yang and W. Li, Orbital-selective superconductivity in the pressurized bilayer nickelate $La_3Ni_2O_7$: An infinite projected entangled-pair state study, Phys. Rev. B **110**, L041111 (2024).
[39] H. Oh and Y. H. Zhang, Type-II *t*-*J* model and shared superexchange coupling from Hund's rule in supercondcuting $La_3Ni_2O_7$, Phys. Rev. B **108**, 174511 (2023).
[40] X. Qu, D. Qu, W. Li and G. Su, Rules of Hund's Rule and Hybridization in the Two-orbital Model for High-$T_c$ Supercondcutivity in the Bilayer Nickelate, arXiv preprint arXiv:2311.12769 (2023).
[41] C. Lu, Z. Pan, F. Yang and C. Wu, Interlayer-Coupling-Driven High-Temperature Superconductivity in $La_3Ni_2O_7$, Phys. Rev. Lett. **132**, 146002 (2024).
[42] Y. F. Yang, G. M. Zhang and F. C. Zhang, Interlayer valence bonds and two-component theory for high-$T_c$ superconductivity of $La_3Ni_2O_7$ under pressure, Phys. Rev. B **108**, L201108 (2023).
[43] J. Li, C. Chen, C. Huang, Y. Han, M. Huo, X. Huang, P. Ma, Z. Qiu, J. Chen, X. Hu, L. Chen, T. Xie, B. Shen, H. Sun, D. Yao and M. Wang, Structural transition, electric transport, and electronic structures in the compressed trilayer nickelate $La_4Ni_3O_{10}$, Sci. China-Phys. Mech. **67**, 117403 (2024).

[44] M. Zhang, C. Pei, X. Du, W. Hu, Y. Cao, Q. Wang, J. Wu, Y. Li, H. Liu, C. Wen, Y. Zhao, C. Li, W. Cao, S. Zhu, Q. Zhang, N. Yu, P. Cheng, L. Zhang, Z. Li, J. Zhao, Y. Chen, H. Guo, C. Wu, F. Yang, S. Yan, L. Yang and Y. Qi, Superconductivity in trilayer nickelate $La_4Ni_3O_{10}$ under pressure, arXiv preprint arXiv:2311.07423 (2023).
[45] Q. Li, Y. Zhang, Z. Xiang, Y. Zhang, X. Zhu and H. Wen, Signature of Superconductivity in Pressurized $La_4Ni_3O_{10}$, Chin. Phys. Lett. **41**, 017401 (2024).
[46] H. Sakakibara, M. Ochi, H. Nagata, Y. Ueki, H. Sakurai, R. Matsumoto, K. Terashima, K. Hirose, H. Ohta, M. Kato, Y. Takano and K. Kuroki, Theoretical analysis on the possibility of supercondcutivity in the trilayer Ruddlesden-Popper nickelate $La_4Ni_3O_{10}$ under pressure and its experimental examination: Comparison with $La_3Ni_2O_7$, Phys. Rev. B **109**, 144511 (2024).
[47] Y. Zhu, D. Peng, E. Zhang, B. Pan, X. Chen, L. Chen, H. Ren, F. Liu, Y. Hao, N. Li, Z. Xing, F. Lan, J. Han, J. Wang, D. Jia, H. Wo, Y. Gu, Y. Gu, L. Ji, W. Wang, H. Gou, Y. Shen, T. Ying, X. Chen, W. Yang, H. Cao, C. Zheng, Q. Zeng, J. Guo and J. Zhao, Superconductivity in pressurized trilayer $La_4Ni_3O_{10-\delta}$ single crystals, Nature **631**, 531 (2024).
[48] X. Du, Y. Li, Y. Cao, C. Pei, M. Zhang, W. Zhao, K. Zhai, R. Xu, Z. Liu, Z. Li, J. Zhao, G. Li, Y. Chen, Y. Qi, H. Guo and L. Yang, Correlated Electronic Structure and Density-Wave Gap in Trilayer Nickelate $La_4Ni_3O_{10}$, arXiv preprint arXiv:2405.19853 (2024).
[49] M. Kakoi, T. Oi, Y. Ohshita, M. Yashima, K. Kuroki, T. Kato, H. Takahashi, S. Ishiwata, Y. Adachi, N. Hatada, T. Uda and H. Mukuda, Multiband Metallic Ground State in Multilayered Nickelates $La_3Ni_2O_7$ and $La_4Ni_3O_{10}$ Probed by $^{139}$La-NMR at Ambient Pressure, J. Phys. Soc. Jpn. **93**, 053702 (2024).
[50] H. Oh, B. Zhou and Y. Zhang, Type II t-J model in charge transfer regime in bilayer $La_3Ni_2O_7$ and trilayer $La_4Ni_3O_{10}$, arXiv preprint arXiv:2405.00092 (2024).
[51] Q. Qin, J. Wang and Y. Yang, Frustrated Superconductivity in the Trilayer Nickelate $La_4Ni_3O_{10}$, arXiv preprint arXiv:2405.04340 (2024).
[52] Y. Zhang, L. Lin, A. Moreo, T. Maier and E. Dagotto, Prediction of $s^{\pm}$-wave supercondcutivity enhanced by electronic doping in trilayer nickelate $La_4Ni_3O_{10}$ under pressure, arXiv preprint arXiv:2402.05285 (2024).
[53] C. Q. Chen, Z. Luo, M. Wang, W. Wu and D. X. Yao, Trilayer multiorbital models of $La_4Ni_3O_{10}$, Phys. Rev. B **110**, 014503 (2024).
[54] Q. G. Yang, K. Y. Jiang, D. Wang, H. Y. Lu and Q. H. Wang, Effective model and $s_{\pm}$-wave superconductivity in trilayer nickelate $La_4Ni_3O_{10}$, Phys. Rev. B **109**, L220506 (2024).
[55] J. X. Wang, Z. Ouyang, R. Q. He and Z. Y. Lu, Non-Fermi liquid and Hund correlation in $La_4Ni_3O_{10}$ under high pressure, Phys. Rev. B **109**, 165140 (2024).
[56] I. V. Leonov, Electronic structure and magnetic correlations in the trilayer nickelate superconductor $La_4Ni_3O_{10}$ under pressure, Phys. Rev. B **109**, 235123 (2024).
[57] P. Blaha, K. Schwarz, F. Tran, R. Laskowski, G. Madsen and L. Marks, WIEN2k: An APW+lo program for calculating the properties of solids, J. Chem. Phys. **152** (2020).

[58] K. Haule, C. Yee and K. Kim, Dynamical mean-filed theory within the full-potential methods: Electronic structure of $CeIrIn_5$, $CeCoIn_5$, and $CeRhIn_5$, Phys. Rev. B **81**, 195107 (2010).
[59] H. LaBollita, J. Kapeghian, M. Norman and A. Botana, Electronic structure and magnetic tendencies of trilayer $La_4Ni_3O_{10}$ under pressure: Structural transition, molecular orbitals, and layer differentiation, Phys. Rev. B **109**, 195151 (2024).
[60] Q. Gu and H. Wen, Superconductivity in nickel-based 112 systems, The Innovation **3**, 100202 (2022).
[61] Z. Yin, K. Haule and G. Kotliar, Kinetic frustration and the nature of the magnetic and paramagnetic states in iron pnictides and iron chalcogenides, Nat. Mater. **10**, 932 (2011).
[62] M. Kakoi, T. Kaneko, H. Sakakibara, M. Ochi and K. Kuroki, Pair correlations of the hybridized orbitals in a ladder model for the bilayer nickelate $La_3Ni_2O_7$, Phys. Rev. B **109**, L201124 (2024).

# Supplemental Materials

# Electronic Correlations and Hund's Rule Coupling in Trilayer Nickelate $La_4Ni_3O_{10}$

Zihao Huo[1], Peng Zhang[2,*], Zihan Zhang[1], Defang Duan[1,*], Tian Cui[3,1]

[1]*Key Laboratory of Material Simulation Methods & Software of Ministry of Education, State Key Laboratory of Superhard Materials, College of Physics, Jilin University, Changchun 130012, China*

[2]*MOE Key Laboratory for Non-equilibrium Synthesis and Modulation of Condensed Matter, Shaanxi Province Key Laboratory of Advanced Functional Materials and Mesoscopic Physics, School of Physics, Xi'an Jiaotong University, Xi'an 710049, China*

[3]*Institute of High Pressure Physics, School of Physical Science and Technology, Ningbo University, Ningbo 315211, China*

*Corresponding author: duandf@jlu.edu.cn, zpantz@mail.xjtu.edu.cn

## I. Computational details

We perform fully charge self-consistent DFT+DMFT calculations using the eDMFT package developed by Haule *et al*. [1] with the Wien2k package [2] for the DFT part. The exchange-correlation functional was described using the Perdew-Burke-Ernzerhof exchange-correlation potential [3] and 11 × 11 × 3 *k*-points were used in both $P2_1/a$- and $I4$/mmm-$La_4Ni_3O_{10}$. The double counting method in DFT+DMFT was treated using the nominal scheme [1], which was suggested as a suit scheme in the study of $La_3Ni_2O_7$ [4]. The hybridization expansion continuous-time quantum Monte Carlo method (CTQMC) [5-7] is used as the impurity solver of the self-consistent DMFT equation [8]. In each DMFT iteration, $6\times10^8$ Monte Carlo updates are used and the self-energy is derived from Dyson's equation. The converged self-energy from DMFT is used to update the new charge density and the new Kohn-Sham potential for the next DFT calculation. The DFT+DMFT loops iterate until the fully convergence of the charge density, the self-energy and the hybridization functions and so on. The maximum entropy method [9] is used for analytic continuation of the self-energy from the imaginary frequency to the real frequency. We choose the on-site Hubbard $U$ = 3.5 eV and the Hund's rule coupling $J$ = 1 eV to parameterize the electron correlations in the Ni-3*d* orbitals. Additionally, several different alternative Hubbard $U$ and Hund's rule coupling $J$ ($U$ = 3.5 eV, $J$ = 0.5 eV and $U$ = 2 eV, $J$ = 1 eV) have been tested, as shown in Fig. S1 and S2.

## II. Supplemental Figures

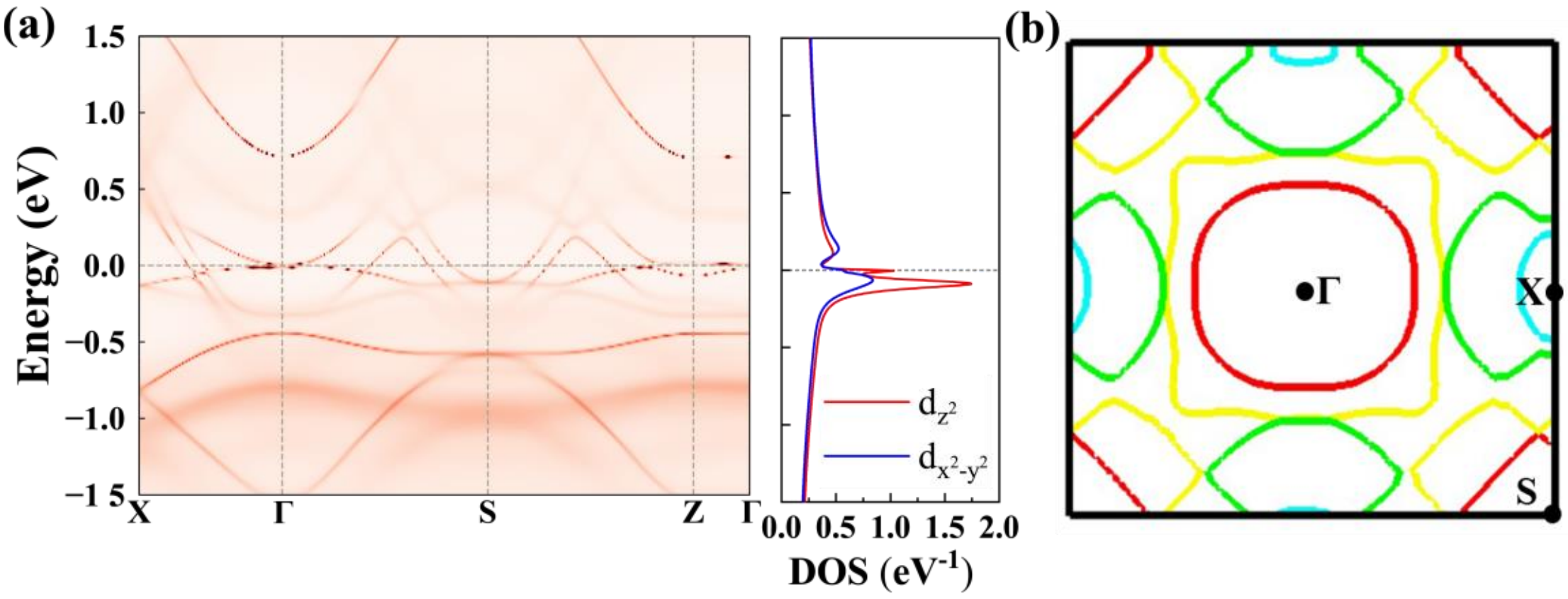

Figure S1: (a) ***k***-resolved spectral functions and DOS of the $P2_1/a$ phase of $La_4Ni_3O_{10}$ calculated by DFT+DMFT ($U$ = 3.5 eV, $J$ = 0.5 eV) at 100 K. The horizontal gray dashed line represents the Fermi level. (b) Two-dimensional Fermi surface of the $P2_1/a$ phase of $La_4Ni_3O_{10}$ calculated by DFT+DMFT at 100 K. Different colors represent different band index.

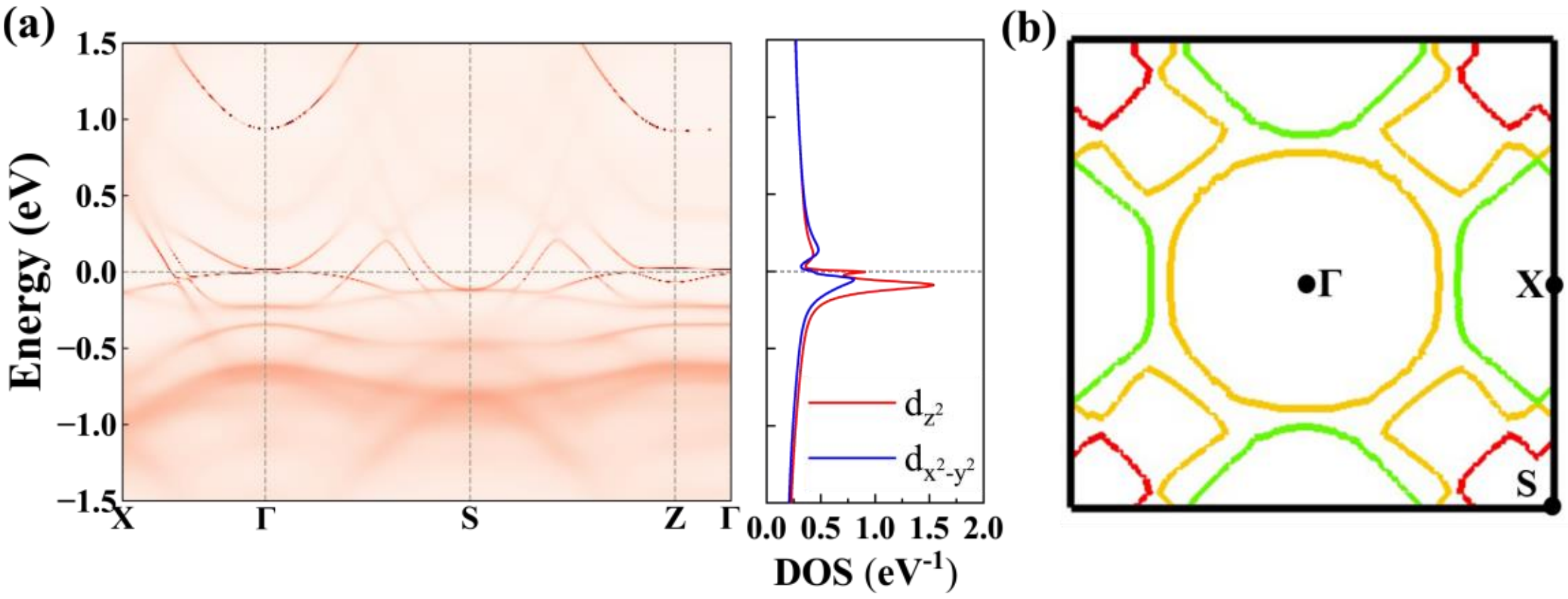

Figure S2. (a) ***k***-resolved spectral functions and DOS of the $P2_1/a$ phase of $La_4Ni_3O_{10}$ calculated by DFT+DMFT ($U$ = 2 eV, $J$ = 1 eV) at 100 K. The horizontal gray dashed line represents the Fermi level. (b) Two-dimensional Fermi surface of the $P2_1/a$ phase of $La_4Ni_3O_{10}$ calculated by DFT+DMFT at 100 K. Different colors represent different band index.

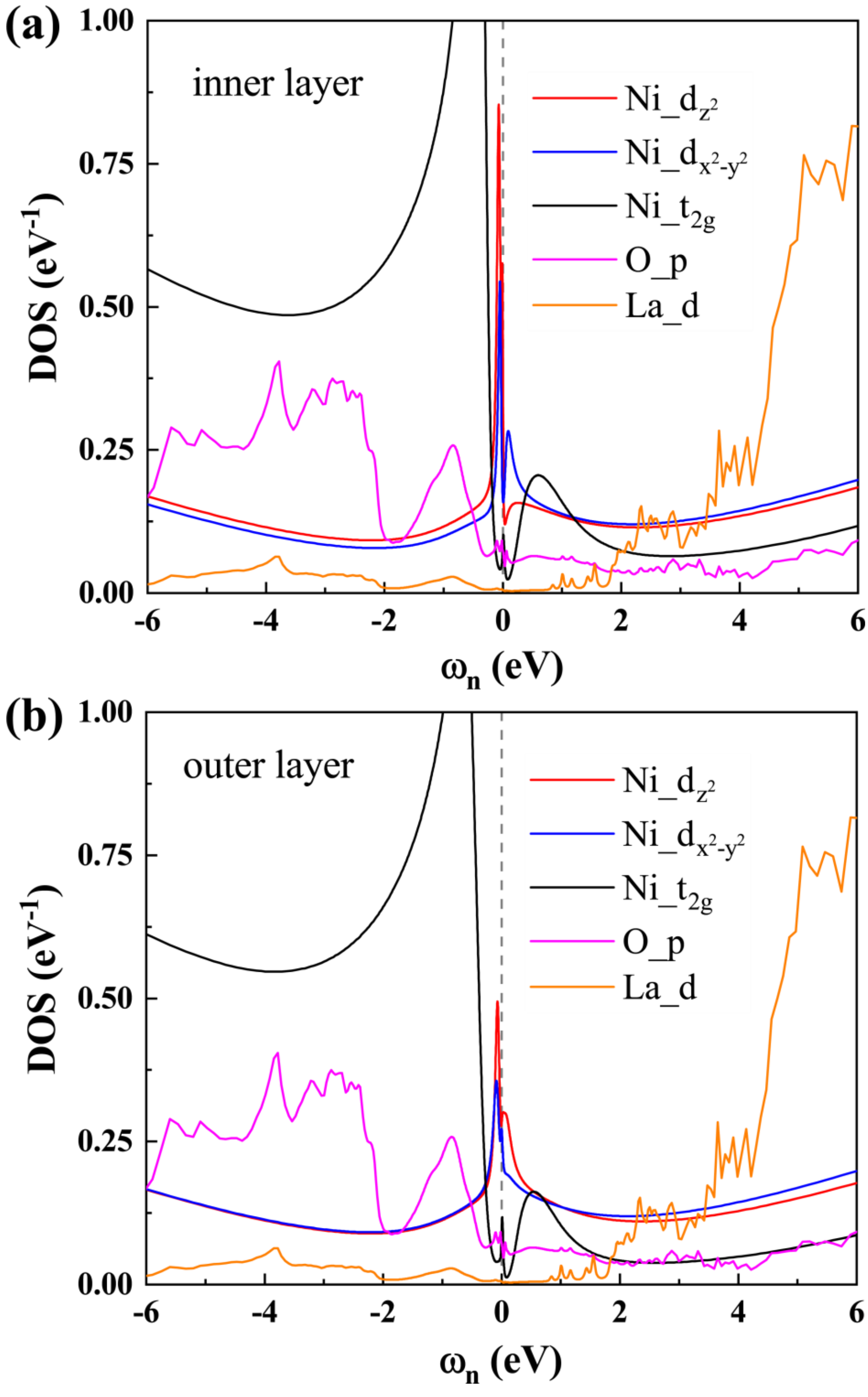

Figure S3. Orbital-dependent electronic DOS of $P2_1/a$ phase of $La_4Ni_3O_{10}$ in (a) inner and (b) outer layer at 100 K. The grey dashed line indicated the Fermi level.

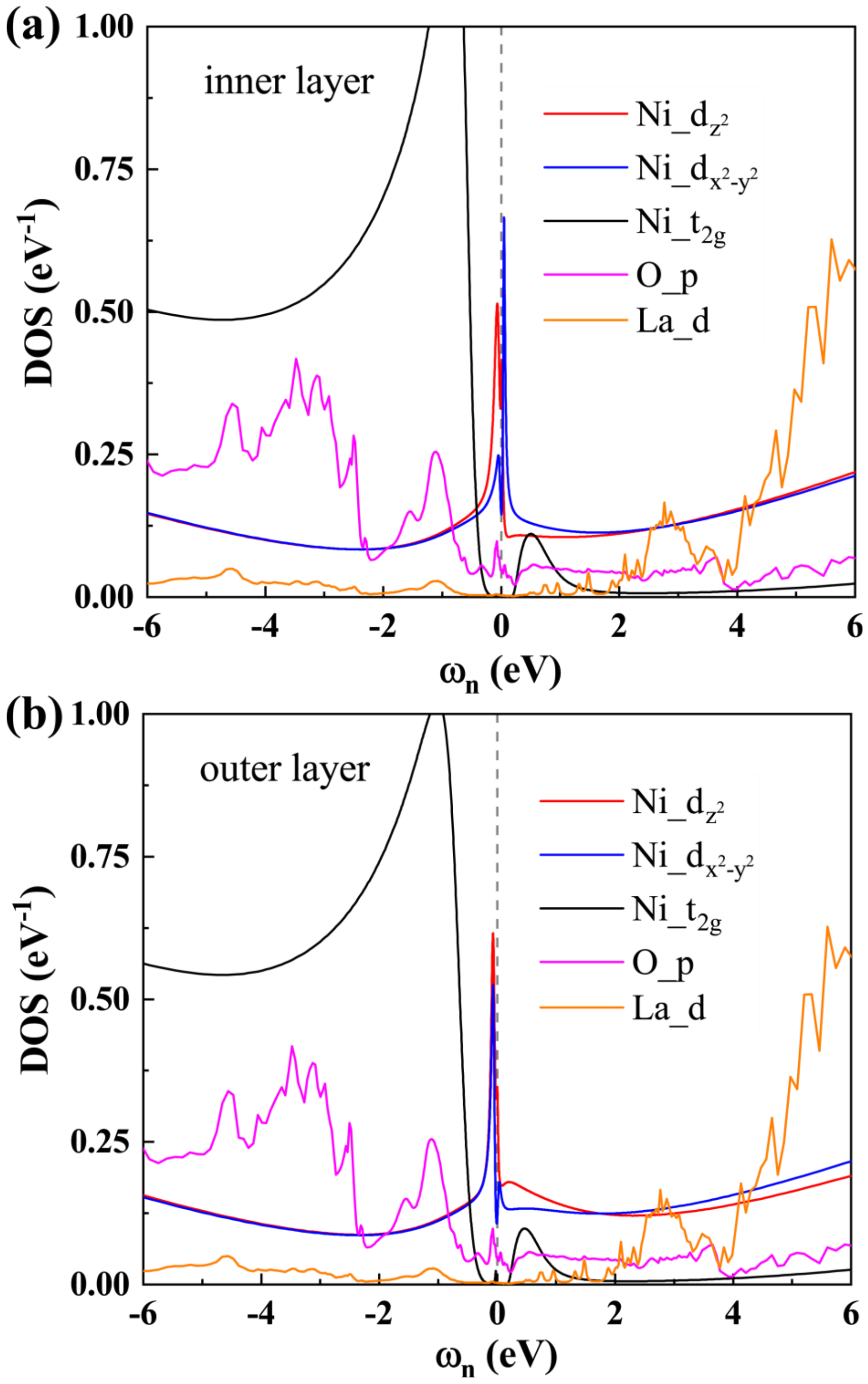

Figure S4. Orbital-dependent electronic DOS of *I*4/*mmm* phase of $La_4Ni_3O_{10}$ in (a) inner and (b) outer layer at 100 K. The grey dashed line indicated the Fermi level.

## III. Supplemental Tables

Table S1. Detailed structural information of RP phase $La_3Ni_2O_7$ and $La_4Ni_3O_{10}$.

| Phase | Pressure (GPa) | Lattice Parameters (Å) | Atomic Coordinates |
|---|---|---|---|
| $P2_1/a$-$La_4Ni_3O_{10}$ | 0 | a=5.416<br>b=5.468<br>c=14.228<br>α=γ=90<br>β=100.752 | O (4e) 0.782 0.983 0.073<br>O (4e) 0.079 0.958 0.645<br>O (4e) 0.119 0.743 0.236<br>O (4e) 0.108 0.245 0.212<br>O (4e) 0.746 0.739 0.486<br>La (4e) 0.805 0.511 0.103<br>La (4e) 0.94 0.499 0.362<br>Ni (4e) 0.857 0.991 0.227<br>Ni (2d) 0.0 0.0 0.5 |
| $I4/mmm$-$La_4Ni_3O_{10}$ | 30.5 | a=b=3.709<br>c=26.716<br>α=β=γ=90 | O (4e) 0.0 0.0 0.071<br>O (4e) 0.0 0.0 0.215<br>O (4c) 0.0 0.5 0.0<br>O (8g) 0.0 0.5 0.138<br>La (4e) 0.0 0.0 0.302<br>La (4e) 0.0 0.0 0.433<br>Ni (2a) 0.0 0.0 0.0<br>Ni (4e) 0.0 0.0 0.14 |
| $I4/mmm$-$La_3Ni_2O_7$ | 30 | a=b=3.657<br>c=19.543<br>α=β=γ=90 | O (4e) 0.0 0.0 0.795<br>O (4e) 0.0 0.0 0.905<br>O (2a) 0.0 0.0 0.0<br>La (4e) 0.0 0.0 0.321<br>La (2b) 0.0 0.0 0.5<br>Ni (4e) 0.0 0.0 0.972 |

## III. References

[1] K. Haule, C. Yee and K. Kim, Dynamical mean-filed theory within the full-potential methods: Electronic structure of $CeIrIn_5$, $CeCoIn_5$, and $CeRhIn_5$, Phys. Rev. B **81**, 195107 (2010).

[2] P. Blaha, K. Schwarz, F. Tran, R. Laskowski, G. Madsen and L. Marks, WIEN2k: An APW+lo program for calculating the properties of solids, J. Chem. Phys. **152** (2020).

[3] J. Perdew, K. Burke and M. Ernzerhof, Generalized Gradient Approximation Made Simple, Phys. Rev. Lett. **77**, 3865 (1996).

[4] Y. Cao and Y. Yang, Flat bands promoted by Hund's rule coupling in the candidate double-layer high-temperature superconductor $La_3Ni_2O_7$ under high pressure, Phys. Rev. B **109**, L081105 (2024).

[5] E. Gull, A. Millis, A. Lichtenstein, A. Rubtsov, M. Troyer and P. Werner, Continuous-time Monte Carlo methods for quantum impurity models, Rev. Mod. Phys. **83**, 349 (2011).

[6] P. Werner, A. Comanac, L. de' Medici, M. Troyer and A. Millis, Continuous-Time Solver for Quantum Impurity Models, Phys. Rev. Lett. **97**, 076405 (2006).

[7] K. Haule, Quantum Monte Carlo impurity solver for cluster dynamical mean-field theory and electronic structure calculations with adjustable cluster base, Phys. Rev. B **75**, 155113 (2007).

[8] A. Georges, G. Kotliar, W. Krauth and M. Rozenberg, Dynamical mean-field theory of strongly correlated fermion systems and the limit of infinite dimensions, Rev. Mod. Phys. **68**, 13 (1996).

[9] M. Jarrell and J. Gubernatis, Bayesian inference and the analytic continuation of imaginary-time quantum Monte Carlo data, Phys. Rep. **269**, 133 (1996).